\documentclass[smallcondensed,final]{svjour3}

\smartqed  

\usepackage{array}
\usepackage{amsmath}
\usepackage{amssymb}
\usepackage{bm}
\usepackage{graphics}
\usepackage{mathrsfs}
\usepackage{verbatim}
\usepackage{subfigure}
\usepackage{placeins}
\usepackage{overpic}
\usepackage[numbers]{natbib}
\usepackage{marvosym}
\usepackage{xcolor}
\usepackage{hyperref}
\usepackage[normalem]{ulem}

\usepackage{subfigure}
\usepackage[ruled,vlined,linesnumbered]{algorithm2e}
\usepackage{booktabs}

\begin{document}

\title{Strong Scaling of Numerical Solver for Supersonic Jet Flow Configurations}

\titlerunning{Strong Scaling of Num. Solver for Supersonic Jet Flow Config.}
\journalname{J. Braz. Soc. Mech. Sci. Eng.}

\author{Carlos Junqueira-Junior \and Jo\~{a}o Luiz F. Azevedo 
\and Jairo Panetta \and William R. Wolf \and Sami Yamouni}

\institute{Carlos Junqueira-Junior (\Letter)	       
      \at \'{E}cole Nationale Sup\'{e}rieure d'Arts et M\'{e}tiers, 
      DynFluid Laboratory, 75013, Paris, France \\ 
	  \email{junior.junqueira@ensam.eu}
	  \and Jo\~{a}o Luiz F. Azevedo 
	  \at Instituto de Aeron\'{a}utica e Espa\c{c}o, DCTA/IAE/ALA, 
	  S\~{a}o Jos\'{e} dos Campos, SP, 12228-904, Brazil
	  \and Jairo Panetta 
	  \at Instituto Tecnol\'{o}gico de Aeron\'{a}utica, DCTA/ITA, 
	  S\~{a}o Jos\'{e} dos Campos, SP, 12228-900, Brazil 
	  \and William R. Wolf
	  \at Faculdade de Engenharia Mec\^{a}nica,
	  Universidade Estadual de Campinas, Rua Mendeleyev, 200,
	  Campinas, SP, 13083-860, Brazil
	  \and Sami Yamouni, \at LatAm Experian DataLab, 
	  S\~{a}o Paulo, SP, Brazil
      }

\date{Received: 2018/11/02 / Accepted: 2019/10/27 / Published: 2019/11/06}

\maketitle

\begin{abstract}

Acoustics loads are rocket design constraints which push 
researches and engineers to invest efforts in the 
aeroacoustics phenomena which is present on launch vehicles. 
Therefore, an in-house computational fluid dynamics tool 
is developed in order to reproduce high fidelity results 
of supersonic jet flows for aeroacoustic analogy 
applications. The solver is written using the large eddy 
simulation formulation that is discretized using a finite 
difference approach and an explicit time integration. 
Numerical simulations of supersonic jet flows are very 
expensive and demand efficient high-performance computing. 
Therefore, non-blocking message passage interface protocols 
and parallel Input/Output features are implemented into 
the code in order to perform simulations which demand up 
to one billion degrees of freedom. The present work 
evaluates the parallel efficiency of the solver when 
running on a supercomputer with a maximum theoretical peak 
of 127.4 TFLOPS. Speedup curves are generated using nine 
different workloads. Moreover, the validation results of 
a realistic flow condition are also presented in the 
current work.
	

\end{abstract}


\section{Introduction}
\label{sec:intro}

One of the main design issues related to launch vehicles 
lies on noise emission originated by the complex 
interaction between the high-temperature/high-velocity 
exhaustion gases and the atmospheric air. These emissions, 
which have high noise levels, can damage the launching 
structure or even be reflected upon the vehicle 
structure itself and the equipment onboard at the top of 
the vehicles. The resulting pressure fluctuations can damage 
the solid structure of different parts of the launcher or 
of the onboard scientific equipment by vibrational acoustic 
stress. Therefore, it is strongly recommended to consider 
the load resulted from acoustic sources over large launching 
vehicles during take-off and also during the transonic flight. 

The authors are interested in studying unsteady property fields
of compressible jet flow configurations in order to eventually 
understand the acoustic phenomena, which are important design 
constraints for rocket applications. Experimental techniques 
used to evaluate such flow configuration are complex and require 
considerably expensive apparatus. Therefore, the authors have 
developed a numerical tool, JAZzY \citep{Junior16}, based on the 
large eddy simulation (LES) formulation \citep{Garnier09} in order 
to perform time-dependent simulations of compressible jet flows. 
JAZzY is a compressible LES parallel code for the calculation 
of supersonic jet flow configurations. The large eddy simulation
approach has been successfully used by the scientific community 
and can provide high fidelity numerical data for aeroacoustic 
applications \citep{Bodony05i8, Wolf_doc, lo2012, wolf2012, 
Junior18-abcm}. The numerical tool is written in the Fortran 90 
standards coupled with Message Passing Interface (MPI) features 
\citep{Dongarra95}. The HDF5 \citep{folk11,folk99} and CGNS 
libraries \citep{cgns2012,legensky02,Poirier00,Poirier98} are 
included into the numerical solver in order to implement a 
hierarchical data format (HDF) and to perform Input/Output 
operations efficiently. 

Large eddy simulations require a significant amount of 
computational resources to provide high fidelity results. The 
scientific community has been using up to hundreds of million 
degrees of freedom on simulations of turbulent flow 
configurations \cite{cohen18,gloerfelt19,Junior18-abcm,
sciacovelli17}. Researchers and engineers need to be certain 
that calculations are run with maximum parallel efficiency when 
allocating computational resources because the access to 
supercomputers is often restricted and limited. Therefore, it 
is of major importance to perform scalability studies, regarding 
an optimal choice of computational load and resources, before 
running simulations on supercomputers. 

The present work addresses the computational performance 
evaluation of the code when using  an Hewlett Packward 
Enterprise (HPE) cluster from a computational center \cite{cepid}. 
The high-performance computing (HPC) solution provides a maximum 
theoretical peak performance of 127.4 TFLOPS using CPUs, Nvidia 
GPUs and Xeon Phi accelerators. Simulations of a pre-defined 
perfectly expanded jet flow condition are performed using 
different loads and resources in order to study the strong 
scalability of the solver. More specifically, nine mesh 
configurations are investigated running on up to 400 processors 
in parallel. The number of degrees of freedom starts with 5.8 
million points and scales to 1.0 billion points. The speedup 
and computational efficiency curves are measured for each grid 
configuration. 

The supercomputer is described after the introduction section. 
Then, numerical formulations and the implementation aspects of the 
tool are discussed. In the sequence, the strong scalability results 
are presented to the reader followed by the discussion on the 
validation of the solver. In the end, the one can find the 
concluding remarks and acknowledgements.




\section{Computational Resources}

The current work is included in the CEPID-CeMEAI \cite{cepid} 
project from the Applied Mathematics department of the 
University of S\~{a}o Paulo. This project is addressed to four 
main research subjects: optimization and operational research; 
computational intelligence and software engineering; computational 
fluid mechanics; and risk evaluation. The CEPID-CeMEAI provides 
access to a high-performance computer server located at the 
University of S\~{a}o Paulo, named Euler. The system presents a 
maximum theoretical peak performance of 127.4 TFLOPS using a 
hybrid parallel processing architecture which has 144 CPU nodes, 
4 fat-nodes, 6 GPU nodes and 1 Xeon Phi node with a total of 3428 
computational cores. The detailed description of each node 
configuration is presented in Tab.\ \ref{tab:computer}. Two 
storage systems are available with 175Tb each one, the network 
file system (NFS) and the {\it Lustre\textsuperscript{\textregistered}} 
file system \cite{lustre}. The network communication is performed 
using Infiniband and Gigabit Ethernet. Red Hat Enterprise Linux 
\cite{redhat} is the operating system of the cluster and Altair 
PBS Pro \cite{pbs} is the job scheduler available. 
\begin{table}[htb!]
\small\sf\centering
\caption{Computer cluster configuration.}\label{tab:computer}
\begin{tabular}{llll}
\hline
Nodes & Processor & Memory & Nb.\ Cores \\
\hline
104 & 2 x CPU {\it Intel\textsuperscript{\textregistered}}
              {\it Xeon\textsuperscript{\textregistered}}
E5-2680v2 & 128GB DDR3 & 20 \\ 
40  & 2 x CPU {\it Intel\textsuperscript{\textregistered}}
              {\it Xeon\textsuperscript{\textregistered}}
E5-2680v4 & 128GB DDR3 & 28 \\ 
4   & 2 x CPU {\it Intel\textsuperscript{\textregistered}}
              {\it Xeon\textsuperscript{\textregistered}}
E5-2667v4 & 512GB DDR3 & 16 \\ 
6   & 2 x CPU {\it Intel\textsuperscript{\textregistered}}
              {\it Xeon\textsuperscript{\textregistered}}
E5-2650v4 & 128GB DDR3 & 24 \\ 
    & + 1 x GPU {\it Nvidia\textsuperscript{\textregistered}}
              {\it Tesla\textsuperscript{\textregistered}}
P-100 & 16GB DDR3 & 3584 \\ 
1   & 2 x CPU {\it Intel\textsuperscript{\textregistered}}
              {\it Xeon\textsuperscript{\textregistered}}
E5-2680v2 & 128GB DDR3 & 20 \\ 
    & + 1 x {\it Intel\textsuperscript{\textregistered}}
          {\it Xeon Phi\textsuperscript{\texttrademark}}
C2108-RP2 & 8GB DDR3 & 60 \\ 
\hline
\end{tabular}
\end{table}




\section{Large Eddy Simulation Formulation}
\label{sec:form}

The numerical simulations of supersonic jet flow configurations are
performed based on the large eddy simulation formulation \cite{Garnier09}. 
This set of equations is based on the principle of scale separation over 
the governing equations used to represent the fluid dynamics, the 
Navier-Stokes formulation. A filtering procedure can be used to describe
the scale separation in a mathematical formalism. The idea is to 
model the small turbulent structures and to calculate the bigger ones. 
Subgrid scale (SGS) closures are added to the filtered equations 
in order to model the effects of small scale turbulent structures.
The Navier-Stokes equations are written in the current work using
the filtering procedure of Vreman \cite{Vreman1995} as 
\begin{equation}
\begin{array}{c}
\displaystyle \frac{\partial \overline{\rho} }{\partial t} + \frac{\partial}{\partial x_{j}} 
\left( \overline{\rho} \widetilde{  u_{j} } \right) = 0 \, \mbox{,}\\
\displaystyle \frac{\partial}{\partial t} \left( \overline{ \rho } \widetilde{ u_{i} } \right) 
+ \frac{\partial}{\partial x_{j}} 
\left( \overline{ \rho } \widetilde{ u_{i} } \widetilde{ u_{j} } \right)
+ \frac{\partial \overline{p}}{\partial x_{i}} 
- \frac{\partial {\tau}_{ij}}{\partial x_{j}}  
+ \frac{1}{3} \frac{\partial}{\partial x_{j}}\left( {\delta}_{ij} \sigma_{kk}\right)
= 0 \, \mbox{,} \\ 
\displaystyle \frac{\partial \overline{e}}{\partial t} 
+ \frac{\partial}{\partial x_{j}} 
\left[ \left( \overline{e} + \overline{p} \right)\widetilde{u_{j}} \right]
- \frac{\partial}{\partial x_{j}}\left({\tau}_{ij} \widetilde{u_{i}} \right)
+ \frac{1}{3} \frac{\partial}{\partial x_{j}}
\left[ \left( \delta_{ij}{\sigma}_{kk} \right) \widetilde{u_{i}} \right]
+ \frac{\partial {q}_{j}}{\partial x_{j}} = 0 \, \mbox{,}
\end{array}
\label{eq:les}
\end{equation}
in which $t$ and $x_{i}$ are independent variables 
representing time and spatial coordinates of a 
Cartesian coordinate system, $\textbf{x}$, respectively. 
The components of the velocity vector, $\textbf{u}$, are 
written as $u_{i}$ and $i=1,2,3$. Density, pressure and 
total energy per unit volume are written as 
$\rho$, $p$, and $e$, respectively. The 
$\left( \overline{\cdot} \right)$ and 
$\left( \tilde{\cdot} \right)$ operators are used in order 
to represent filtered and Favre averaged properties, 
respectively. 

It is important to remark that the System I filtering 
procedure \cite{Vreman1995} neglects the double correlation 
term, $\widetilde{{u}_{i}{u}_{j}}$, which is present 
into the total energy per unit volume equation
in order to write
\begin{equation}
	\overline{e} = \frac{\overline{p}}{\gamma - 1} 
	+ \frac{1}{2} \rho \widetilde{u}_{i} \widetilde{u}_{i} \, \mbox{.} 
\end{equation}
The heat flux, $q_{j}$, is given by
\begin{equation}
	{q}_{j} = \left(\kappa+{\kappa}_{sgs}\right) 
	\frac{\partial \widetilde{T}}{\partial x_{j}} 
	\, \mbox{.}
	\label{eq:q_mod}
\end{equation}
where $T$ and $\kappa$ stand for static temperature and the 
thermal conductivity coefficient, respectively. The last can 
be expressed as
\begin{equation}
\kappa = \frac{\mu C_{p}}{Pr} \, \mbox{,}
\end{equation}
in which $Cp$ is the specific heat at constant pressure, 
$\mu$ is the dynamic viscosity coefficient and $Pr$ is the Prandtl number, 
which is equal to $0.72$ for air. The SGS thermal conductivity coefficient, 
$\kappa_{sgs}$, is written as
\begin{equation}
	\kappa_{sgs} = \frac{\mu_{sgs} C_{p}}{ {Pr}_{sgs} } 
	\, \mbox{,}
	\label{eq:kappa_sgs}
\end{equation}
where ${Pr}_{sgs}$ is the SGS Prandtl number, which is
equal to $0.9$ for static SGS closures and $\mu_{sgs}$
is the eddy viscosity coefficient that is calculated by the SGS
model. In the present work, the dynamic viscosity coefficient 
is calculated using the Sutherland Law which is given by
\begin{eqnarray}
	\mu \left( \widetilde{T} \right) = \mu_{\infty} 
	\left( \frac{\widetilde{T}}{\widetilde{T}_{\infty}}
	\right)^{\frac{3}{2}} 
	\frac{\widetilde{T}_{0}+S_{1}}{\widetilde{T}+S_{1}} \mbox{ , }
	& \mbox{with} \: S_{1} = 110.4K \, \mbox{.}
\label{eq:sutherland}
\end{eqnarray}
One can use an equation of state to correlate density, static 
pressure and static temperature, 
\begin{equation}
	\overline{p} = \overline{\rho} R \widetilde{T} \, \mbox{,}
\end{equation}
in which $R$ is the gas constant, written as
\begin{equation}
R = C_{p} - C_{v} \, \mbox{,}
\end{equation}
and $C_{v}$ is the specific heat at constant volume.
The shear-stress tensor, $\tau_{ij}$, is written 
as,
\begin{equation}
	{\tau}_{ij} = 2 \left(\mu+{\mu}_{sgs}\right) 
	\left( \tilde{S}_{ij} - \frac{1}{3} \delta_{ij} \tilde{S}_{kk} \right) \,
	\label{eq:tau_mod}
\end{equation}
where the components of the rate-of-strain tensor, 
$\tilde{S}_{ij}$, are given by
\begin{equation}
	\tilde{S}_{ij} = \frac{1}{2} 
	\left( \frac{\partial \tilde{u}_{i}}{\partial x_{j}} 
	+ \frac{\partial \tilde{u}_{j}}{\partial x_{i}} 
\right) \, \mbox{.}
\end{equation}
The SGS stress tensor components can be written using the eddy 
viscosity coefficient \cite{Sagaut05},
\begin{equation}
    \sigma_{ij} = - 2 \mu_{sgs} \left( \tilde{S}_{ij} 
	- \frac{1}{3} \tilde{S}_{kk} \right)
    + \frac{1}{3} \delta_{ij} \sigma_{kk}
    \, \mbox{.}
    \label{eq:sgs_visc}
\end{equation}
In the present article, the eddy viscosity coefficient, $\mu_{sgs}$, 
and the components of the isotropic part of the SGS stress tensor, 
$\sigma_{kk}$, are not considered for the calculations. Previous
validation results performed by the authors \cite{Junior16,Junior18-abcm} 
indicate that the characteristics of numerical discretization of JAZzY 
can overcome the effects of subgrid scale models. The same conclusion 
can be found in the work of Li and Wang \cite{Li15}. 
Therefore, one can write the large eddy simulation set of equations 
can be written in a more compact form as
\begin{equation}
  \frac{\partial \textbf{Q}}{\partial t} = -\textbf{RHS} \, \mbox{,}
  \label{eq:vec-les}
\end{equation}
where $\textbf{Q}$ stands for the convervative properties vector,
given by
\begin{equation}
  \textbf{Q} = \left[ 
  \overline{\rho} \, \mbox{,} \,
  \overline{ \rho } \widetilde{ u_{i} } \,\mbox{,} \, 
  \overline{e}
  \right]^{T} \, \mbox{,}
  \label{eq:con-prop}
\end{equation}
and $\textbf{RHS}$, which stands for the right-hand 
side of equation, represents the contribution of inviscid and viscous
fluxes terms from Eq.\ \ref{eq:les}\@. The components of the right-hand
side vector are written as
\begin{equation}
  {RHS}_{i} = 
  \frac{\partial {E}_{i}}{\partial x_{j}} -
  \frac{\partial {F}_{i}}{\partial x_{j}} \, \mbox{,}
  \label{eq:rhs}
\end{equation}
in which $E_{i}$ and $F_{i}$, are the components of inviscid and 
viscous flux vectors, respectively given by 
\begin{equation}
  \begin{array}{ccc}
    \textbf{E}=\left[
      \begin{array}{c}
        \overline{\rho} \widetilde{u_{j}}\\
  	    \overline{\rho} \widetilde{u_{i}} \widetilde{u_{j}} +
        \overline{p}\delta_{ij} \\
        \left[ \left( \overline{e} + \overline{p} \right)
        \widetilde{u_{j}} \right]
      \end{array}
    \right] & 
    \mbox{and} &
    \textbf{F}=\left[
      \begin{array}{c}
        0 \\
  	    {\tau}_{ij}\\
        {\tau}_{ij} \widetilde{u_{i}} - {q}_{j}
      \end{array}
    \right] \, \mbox{.} 
  \end{array}
\end{equation}

Spatial derivatives are calculated in a 
structured finite difference context and the formulation 
is re-written for the general curvilinear coordinate system 
\cite{Junior16}. The numerical flux is computed through 
a central difference scheme with the explicit addition 
of anisotropic scalar artificial dissipation model of 
Turkel and Vatsa \cite{Turkel_Vatsa_1994}. The time 
marching method is an explicit 5-stage Runge-Kutta scheme  
developed by Jameson {\em et. al.} \cite{Jameson81}. 

\begin{figure}[htb]
  \begin{center}
    \includegraphics[width=0.95\textwidth,trim=0.25cm 0.25cm 0.25cm 0.25cm,clip]
    {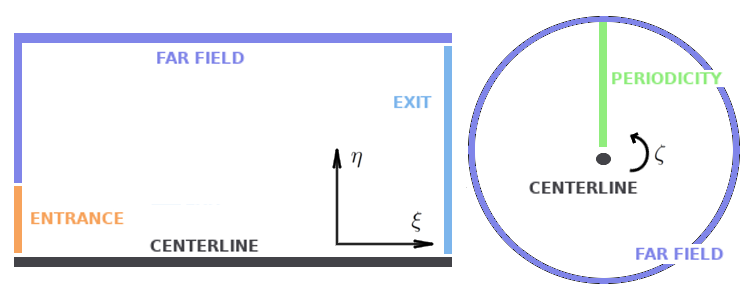} 
	\caption{Two-dimensional lateral and frontal illustrations of the 
    domain which indicate the positioning of boundary conditions.}
    \label{fig:bc}
  \end{center}
\end{figure}

Boundary conditions for the LES formulation are imposed in 
order to represent a supersonic jet flow into a 3-D 
computational domain with cylindrical shape. 
Figure \ref{fig:bc} presents a lateral view and a frontal 
view of the computational domain used in the present work 
and the positioning of the entrance, exit, centerline, 
far field, and periodic boundary conditions.

A flat-hat velocity profile is implemented at the entrance 
boundary condition through the use of the 1-D characteristic 
relations for the 3-D Euler equations in order to create a 
jet-like flow configuration. The set of properties, then, 
determined is computed from within and from outside the 
computational domain. Riemann invariants \cite{Long91} are 
used in order to calculate properties at the far field 
surfaces where the normal-to-face components of the velocity 
are computed by a zero-order extrapolation from inside the 
computational domain. The angle of flow at the far field 
boundary is assumed fixed. The remaining properties are 
obtained as a function of the jet Mach number, which is a
known variable. 

The flow configuration is assumed to be subsonic at the exit 
plane of the domain. Therefore, the pressure is obtained from 
the outside, {\em i.e.}, it is assumed given, the internal 
energy and components of the velocity are calculated by 
a zero-order extrapolation from the interior of the domain. 
Then, density, $\rho$, and total energy per unit volume, $e$,
are computed at the exit boundary using the extrapolated 
properties and the imposed pressure at the output plane.
The first and last points in the azimuthal direction are 
superposed in order to close the 3-D computation domain and 
create a periodicity boundary condition.
An adequate treatment of the centerline boundary is necessary
since it is a singularity of the coordinate transformation.
The conserved properties are extrapolated from the adjacent
longitudinal plane and averaged in the azimuthal direction in 
order to define the updated properties at the centerline of 
the jet. Furthermore, the fourth-difference terms of the 
artificial dissipation scheme of Turkel and Vatsa 
\cite{Turkel_Vatsa_1994} are carefully treated in order to 
avoid five-point difference stencils at the centerline 
singularity. 

The reader can find further details about the spatial 
discretization, time marching scheme and implementation of 
boundary conditions in the work of {\em Junqueira-Junior} 
\cite{Junior16} and {\em Junqueira-Junior et. al.} 
\cite{Junior18-abcm} which present the validation of the 
large eddy simulation solver.




\section{Parallel Implementation Aspects}
\label{sec:impl}

The solver is developed to calculate the LES set of equations, Eq.\ 
\ref{eq:vec-les}, for supersonic jet flow configurations using the 
Fortran 90 standard. The spatial discretization of the formulation 
is based on a centered finite-difference approach with the explicit
addition of anisotropic scalar artificial dissipation model of 
Turkel and Vatsa \cite{Turkel_Vatsa_1994} and the time integration 
is performed using an explicit 2nd-order 5-stage Runge-Kutta scheme 
\cite{Jameson81}. 

Parallelism is achieved using the single program multiple data, SPMD, 
approach \cite{Darema01} and the exchange of messages provided by MPI 
protocols \cite{Dongarra95}. The algorithm of the LES solver is 
structured in two main steps. Firstly, a pre-processing routine reads 
a mesh file and performs a balanced partitioning of the domain 
procedure. Then, in the processing routine, each MPI rank reads its 
correspondent grid file and starts the calculations.

\begin{figure}[htb]
  \begin{center}
    \subfigure[Partitioning of computational domain.]{
    \includegraphics[width=0.475\textwidth]{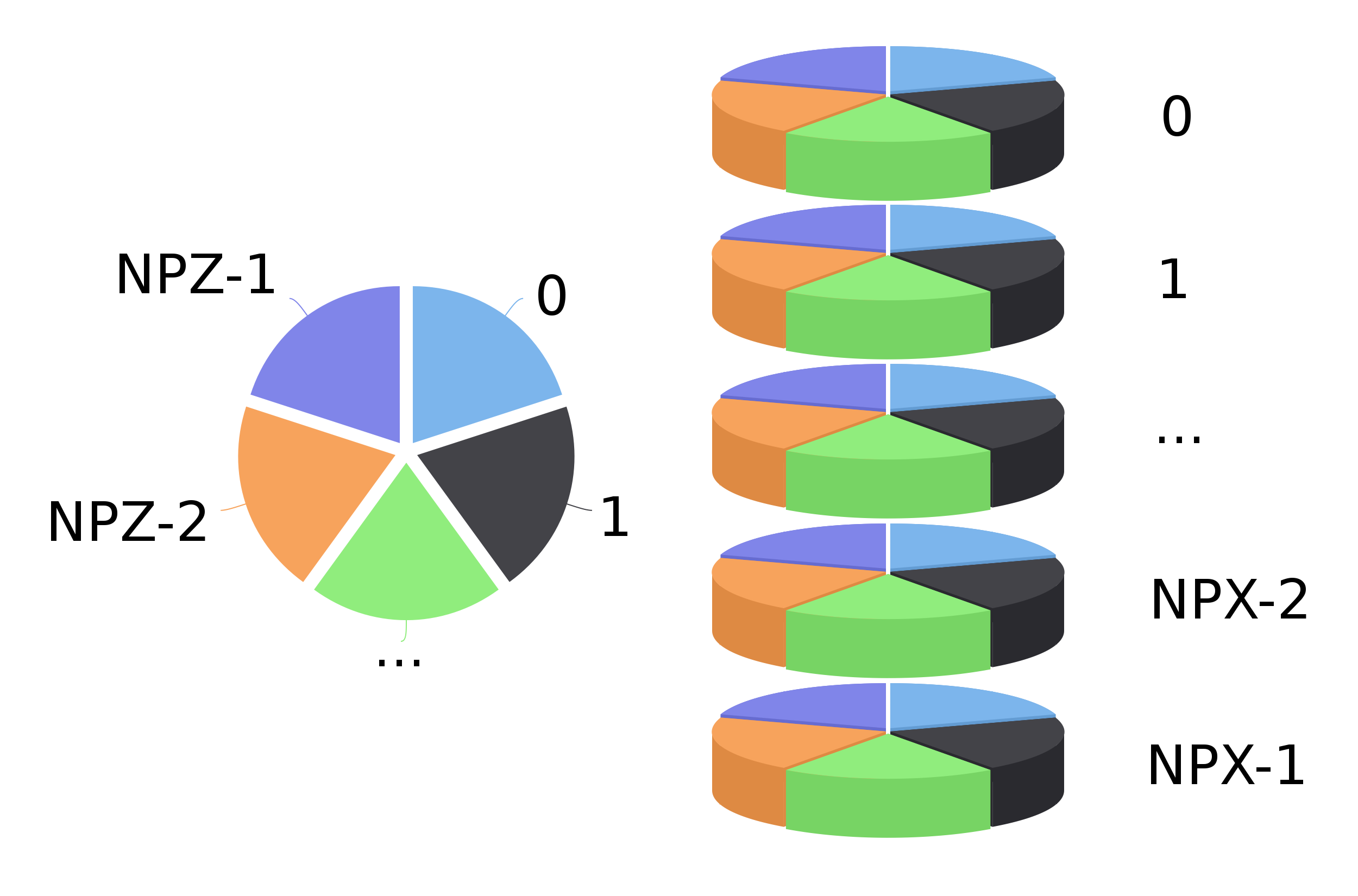}
       \label{fig:part}
    }
	\subfigure[Balancing procedure in one dimension.]{
    \includegraphics[width=0.475\textwidth]{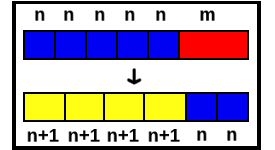}
    \label{fig:balance}
    }
    \caption{Partitioning and balancing approaches.}\label{fig:div-dom}
  \end{center}
\end{figure}

The pre-processing routine is run separately from the processing step. 
It reads an input file with the partitioning configuration and a 2-D 
grid file. Next, the pre-processing code calculates the number of 
points in the axial and azimuthal directions in order to perform the 
partitioning and the extrusion in the 3rd direction for each sub-domain. 
The segmentation of the grid points is illustrated in Figure 
\ref{fig:part}. A matrix index notation is used in order to represent 
positions of each partition where NPX and NPZ denote the number of 
partitions in the axial and azimuthal directions, respectively. 
In the case of a non-exact domain division, the remaining points 
are spread among the partitions in order to have a well-balanced 
task distribution. Algorithm \ref{alg:part} presents the details 
of this division and Fig.\ \ref{fig:balance} illustrates 
the balancing procedure in one direction, where {\em LocNbPt},
{\em TotNbPt}, {\em NbPart}, and {\em PartIndex} stand for local number 
of points in one direction, total number of points in one direction,
number of partitions in one direction, and the index of the partitions
in one direction. The same algorithm is used to perform the 
partitioning procedure in both axial and azimuthal directions. 
This preprocessing part of code is executed sequentially.
\begin{algorithm}[htb]
  \SetKwData{LocNbPt}{LocNbPt}
  \SetKwData{TotNbPt}{TotNbPt}
  \SetKwData{NbPart}{NbPart}
  \SetKwData{PartIndex}{PartIndex}
  \SetKwFunction{mod}{mod}
  \Begin{
	$\textbf{int}$ \LocNbPt\tcp*[h]{Loc. nb. of pt. in one dir.}\;
	$\textbf{int}$ \TotNbPt\tcp*[h]{Tot. nb. of pt. in one dir.}\;
	$\textbf{int}$ \NbPart\tcp*[h]{Nb. of part. in one dir.}\;
	$\textbf{int}$ \PartIndex\tcp*[h]{Part. index in one dir.}\;
	\LocNbPt = $ \left( \TotNbPt / \NbPart \right) $ \;
	\If{$\left[\PartIndex < \mod{\TotNbPt,\NbPart} \right]$}{
	  \LocNbPt = \LocNbPt + 1 \;
    }
  }
  \caption{Optimized partitioning approach.}\label{alg:part}
\end{algorithm}

The mesh files, for each domain, are written after 
the optimized partitioning procedure using the CFD General Notation 
System (CGNS) standard \citep{legensky02,Poirier98,Poirier00,cgns2012}. 
This standard is based on the HDF5 \citep{folk99,folk11} libraries 
which provide tools for hierarchical data format (HDF) and can perform 
Input/Output operations efficiently. The authors have chosen to write 
one CGNS grid file for each partition in order to have each MPI rank 
performing I/O operations independently, {\em i.e.} in parallel, during
the processing step of the calculations. Moreover, each MPI rank can 
also write its own time-dependent solution to a local CGNS file. Such 
an approach avoids synchronizations during check-points, which can 
be a significant drawback in HPC applications.

\begin{figure}[htb!]
  \begin{center}
    \includegraphics[width=0.75\textwidth]{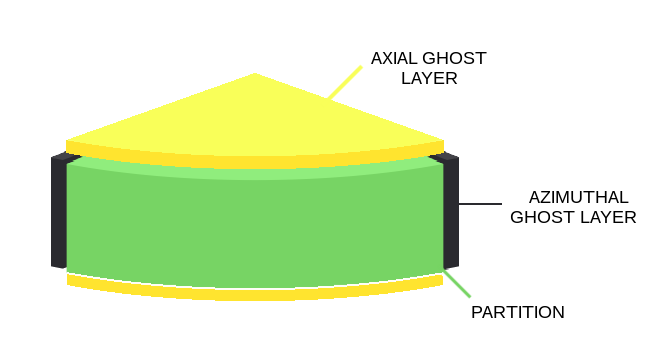}
    \caption{The positioning of ghost layer points.}\label{fig:ghost}
  \end{center}
\end{figure}

After the pre-processing routine, the solver can start the
simulation. A brief overview of the computing part of the 
LES code is presented in Alg.\ \ref{alg:les}.\@ 
\begin{algorithm}[htb]
  \SetKwData{Jac}{Jacob.}\SetKwData{Met}{Metric}
  \SetKwData{IC}{I.C.}\SetKwData{Cons}{Cons.\ Vec.}
  \SetKwData{BC}{B.C.}\SetKwData{Visc}{Visc.}
  \SetKwData{Input}{Input}\SetKwData{Output}{Output}
  \SetKwData{Rest}{Rest.\ data}
  \SetKwData{Mesh}{Mesh}\SetKwData{ArtDiss}{Art.\ Diss.}
  \SetKwData{InvVec}{Inv.\ Flux}\SetKwData{ViscVec}{Visc.\ Flux}
  \SetKwData{RHS}{RHS Vec.}
  \SetKwData{Ghost}{Ghost pts.}
  \SetKwFunction{Read}{Read}\SetKwFunction{Write}{Write}
  \SetKwFunction{Com}{Comm}\SetKwFunction{Upt}{Update}
  \SetKwFunction{Calc}{Calc}\SetKwFunction{Part}{Part}
  \SetKwFunction{Create}{Create}\SetKwFunction{Barr}{MPI Sync.}
  \Begin{
	\Read{\Input} \tcp*{Read flow conf.}
    \Read{\Mesh}  \tcp*{Read local mesh}
    \Calc{\Jac,\Met} \;
    \Create{\Ghost} \;
    \eIf{Restart}{
      \Read{\Rest}\tcp*{Read check-point sol.}
    }{
      \Calc{\IC} \tcp*{Impose I.C.}
    }
    \Com{\Jac,\Met,\IC}\;
	\While(\tcp*[f]{Main It. loop}){Nb. it.}{
      \For(\tcp*[f]{5-steps Runge-Kutta}){$\ell\leftarrow 1$ \KwTo $5$}{
        \Barr \tcp*{MPI Barrier Func.}	
        \For(\tcp*[f]{3-D spatial loop}){i,j,k}{
          \Calc{\InvVec}\tcp*{Calc. of inviscid vec.}
          \Calc{\ArtDiss}\tcp*{Calc. of artficial dissip. terms}
          \Calc{\ViscVec}\tcp*{Calc. of viscous vec.}
          \Calc{\RHS}\tcp*{Calc. of RHS vec.}
          \Upt{\Cons,$\alpha_{\ell}$}\tcp*{$\ell$-th R-K step}
        }
        \Upt{\BC,\Visc}\tcp*{Update B.C. and viscosity coef.}
        \Com{\Cons.}\tcp*{Asynchronous MPI comm.}
      }
	  \Write{\Output}\tcp*{Writes time-dependent CGNS sol.}
    }
  }
  \caption{Implementation of large eddy simulation formulation.}\label{alg:les}
\end{algorithm}
Primarily, every MPI process reads the same ASCII file with 
input data such as flow configurations and simulation settings,
as indicated in line 2 of Alg.\ \ref{alg:les}. In the sequence, 
lines 3 and 4 of the same algorithm, each rank reads a local-domain 
CGNS file, calculates Jacobian and metric terms, which are used 
for the general curvilinear coordinates transformation. Ghost points 
are added to the boundaries of local mesh at the axial and azimuthal 
directions in order to carry information of neighbor partition points,
line 5 in Alg.\ \ref{alg:les}. The artificial dissipation scheme 
of Turkel and Vatsa \cite{Turkel_Vatsa_1994} implemented in 
the code \cite{jameson_mavriplis_86} uses a five points stencil 
which requires information of the two neighbors of a given mesh 
point. Hence, two-layer ghost points are created at the beginning 
and at the end of each partition. Figure \ref{fig:ghost} presents the 
layer of ghost points used in the present code. The yellow and black 
layers represent the axial and azimuthal ghost points respectively. 
The green region represents the local partition grid points.

The initial conditions of the flow configurations are imposed 
following the sequence of tasks of the processing algorithm.
They are calculated using data from a previous checkpoint or, 
from the input data depending on if it is a restart simulation 
or not, as presented in lines 5 to 9 of Alg.\ \ref{alg:les}. 
An asynchronous MPI communication of metric terms, Jacobian 
terms, and conservative properties, set as initial conditions, 
is performed between partitions which share neighbor surfaces 
in order to update all ghost points before starting the time 
integration.

The core of the code consists of the while loop indicated in 
line 11 of Alg.\ \ref{alg:les}. This specific part of the code 
is evaluated in the scalability study presented in Sec.\
\ref{sec:comp-perfo}, {\it Computational Performance Study}. 
This loop performs the Runge-Kutta time integration iteratively, 
for all grid points of the computational domain, until reaching 
the requested number of iterations. A succession of computing 
subroutines is performed at each call for the time marching 
scheme. It starts with synchronization of MPI ranks in order
to avoid race conditions followed by the calculation of the 
inviscid flux vectors, artificial dissipation terms, viscous 
flux vector and right-hand side vector, chronologically. 
The boundary condition and the dynamic viscosity coefficient
are updated at the end of each time-marching step followed
by a nonblocking communication of the conservative properties
at the partition boundaries. If necessary, a time-dependent 
solution is incremented to the CGNS output le at the end of 
the main loop iteration.




\section{Computational Performance Study}\label{sec:comp-perfo}

The numerical discretization approach used in the present 
article requires very refined grids in order to reproduce 
high fidelity results for the simulation of supersonic jet 
flows configurations. Therefore, parallel computing with 
efficient inter-partition data exchanges is mandatory.
The parallel efficiency of the code is measured using 
different computational loads and the results are presented
in the present section. The calculations performed in the 
current article are run using the 104 nodes of the Euler
supercomputer with the 
{\it Intel\textsuperscript{\textregistered} 
Xeon\textsuperscript{\textregistered}} E5-2680v2 
architecture and 128GB DDR3 rapid access memory.

The {\it Intel\textsuperscript{\textregistered}} Composer 
XE compiler, version 15.0.2.164, is used in the present work. 
A set of compiling flags which have been tested in previous 
work \cite{Junior16} are used in the present paper:
$$
 \mbox{--O3 --xHost --ipo --no-prec-div --assume-buffered 
 --override-limits}
$$
where {\bf O3} enables aggressive optimization such as global 
code scheduling, software pipelining, predication and 
speculation, prefetching, scalar replacement and loop 
transformations; {\bf xHost} tells the compiler to generate 
instructions for the highest instruction set available on the 
compilation host processor; {\bf ipo} uses automatic, multi-step 
process that allows the compiler to analyze the code and 
determine where you can benefit from specific optimizations; 
{\bf no-prec-div} enables optimizations that give slightly less 
precise results than full division; {\bf assume-buffered\_io} 
tells the compiler to accumulate records in a buffer; and 
{\bf override-limits} deals with very large, complex functions 
and loops. 


\subsection{Scalability setup}\label{subsec:setup}

Simulations of an unheated perfectly expanded jet flow are 
performed using different grid sizes and number of processors 
in order to study the strong scalability of JAZzY. The jet 
entrance Mach number is $1.4$\@. The pressure ratio, 
$PR=P_{j}/P_\infty$, and the temperature ratio, 
$TR=T_{j}/T_\infty$, between the jet entrance and the ambient 
freestream conditions, are equal to one, {\em i.e.}, $PR = 1$ 
and $TR=1$ where the $j$ subscript identifies the properties at 
the jet entrance and the $\infty$ subscript stands for properties 
at the farfield region. The Reynolds number of the jet is 
$Re = 1.57 \times 10^{6}$, based on the jet entrance diameter, 
D\@. The time increment, $\Delta t$, used for the validation 
study is $1\times 10^{-4}$ dimensionless time units. 
A stagnated flow is used as the initial condition for the 
simulations.

\begin{figure}[htb!]
  \begin{center}
    \includegraphics[width=0.55\textwidth,
	trim=0.25cm 0.25cm 0.25cm 0.25cm,clip]{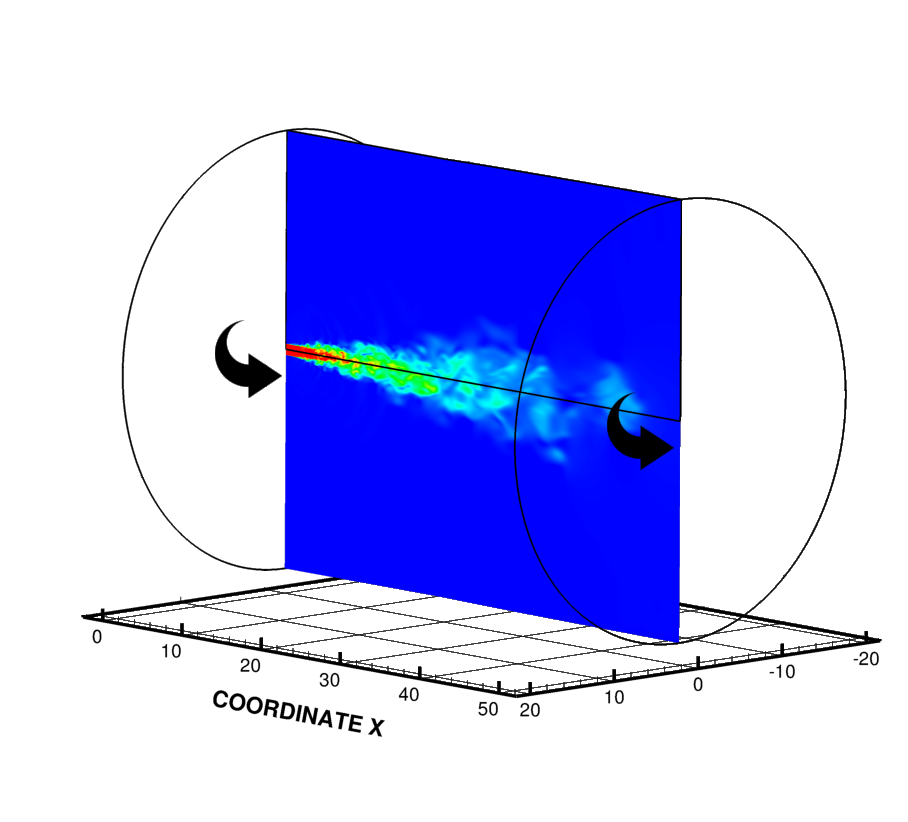} 
	\caption{A Two-dimensional surface, colored by the velocity 
    magnitude contours, extracted from the complete geometry used on the 
    strong scalability study.}\label{fig:geom}
  \end{center}
\end{figure}

The same geometry is used for the computational evaluation, 
where, the 2-D surface of this computational domain, as 
presented in Fig.\ \ref{fig:bc}, is 30 dimensionless length 
and 10 dimensionless height. Figure \ref{fig:geom} illustrates 
a 2-D cut of the geometry coloured by velocity contours.

The present work uses nine different mesh configurations 
whose total number of points doubles every time. Table 
\ref{tab:mesh} presents the details of each grid design
where the first column presents the name of the mesh while 
the second and third columns present the number of points 
in the axial and radial directions, respectively. The last 
column indicates the total number of points of the mesh. 
The grid point distribution in the azimuthal direction is 
fixed at 361. The smallest grid, named as Mesh A, has 
approximately 5.9 million points while the biggest grid 
presents approximately 1.0 billion points. 
\begin{table}[htb!]
\begin{center}
\caption{Configuration of computational meshes used in the
	current study.}
\label{tab:mesh}
\begin{tabular}{cccc}
\hline
Mesh & No.\ Pt.\ Axial Dir.\ & No.\ Pt.\ Radial Dir.\ & 
No.\ Pt. \\
\hline 
	A & 128  & 128  & 5.9M \\
    B & 256  & 128  & 11.8M \\
	C & 256  & 256  & 23.7M \\
	D & 512  & 256  & 47.3M \\
	E & 512  & 512  & 94.6M \\ 
	F & 1024 & 512  & 189.3M \\
	G & 1024 & 1024 & 378.5M \\
	H & 2048 & 1024 & 757.1M \\
	I & 1700 & 1700 & 1.0B \\
\hline
\end{tabular}
\end{center}
\end{table}

The solver is able to have different partitioning configurations 
for a fixed number of sub-domains since the division of the mesh 
is performed in the axial and azimuthal directions. Therefore, 
different partitioning configurations are evaluated for a given 
number of processors. Each simulation performs 1000 iterations or 
24 hours of computation and the average of the CPU time per 
iteration of the while loop indicated in Alg.\ \ref{alg:les} is 
measured in order to calculate the speedups of the solver at the 
Euler supercomputer. The partitioning configurations which provides 
the fastest calculation is used to evaluate the performance of the 
solver. Table \ref{tab:part} presents the number of partitions in 
the azimuthal direction for each number of processors used to study 
the scalability of the solver. The first column stands for the number 
of computational cores while the second column represents the number 
of zones in the azimuthal direction used to evaluate the effects of 
the partitioning on the computation. Calculations performed in the 
present study are run using one single core up to 400 MPI processes. 
\begin{table}[htb!]
\begin{center}
\caption{Number of partitions in the azimuthal direction
	for a given number of processors.}
\label{tab:part}
\begin{tabular}{*9c}
\hline
Nb.\ Proc.\ & \multicolumn{8}{c}{Nb.\ of Part.\ in the 
Azimuthal Dir.\ }  \\
\hline 
	  1   & 1 & &  & & & & & \\
	  2   & 1 & 2  & & & & & & \\
	  5   & 1 & 5  & & & & & & \\
	  10  & 1 & 2  & 5  & 10 & & & & \\
	  20  & 1 & 2  & 4  & 5  & 10 & 20 & & \\
	  40  & 1 & 2  & 4  & 5  & 8  & 10 & 20 & 40 \\
	  80  & 2 & 4  & 5  & 8  & 10 & 20 & 40 & \\
	  100 & 2 & 4  & 5  & 10 & 20 & 25 & & \\
	  200 & 4 & 8  & 10 & 20 & 25 & 50 & & \\
	  400 & 8 & 16 & 20 & 25 & 50 & & & \\
\hline
\end{tabular}
\end{center}
\end{table}

A sequential computation is the ideal starting point, $s$, 
for a strong scalability study. However, frequently, the 
computational problem cannot be allocated in one single 
cluster node due to hardware limitations of the system. 
Therefore, it is necessary to shift the starting point to 
a minimum number of resources in which the code can be run. 
The LES solver has shown to be capable to allocate the five 
smallest grids, from mesh A to mesh E, in one single node 
of the Euler computer. Aforementioned indicates that it is 
possible to run a simulation with 94.6 million grid points 
using 128GB of RAM. The starting points of mesh F and mesh 
G are 40 cores, allocated in two nodes, and 80 cores, 
allocated in four nodes, respectively. Meshes H and I start 
the strong scalability study using 200 cores in 10 nodes of 
the Euler computer. Table \ref{tab:strong} presents the 
minimum number of computational cores used by each mesh 
configuration for the strong scaling study.
\begin{table}[htb]
\small\sf\centering
\caption{Scalability starting point for each mesh 
configuration.}\label{tab:strong}
\begin{tabular}{ll}
\hline
Mesh & $s$\\
\hline
A-E & 1   \\ 
F   & 40  \\ 
G   & 80  \\
H-I & 200 \\
\hline
\end{tabular}
\end{table}



\subsection{Strong scalability study}

The speedup, ${Sp}$, is used in the present work to measure 
the strong scaling of the solver and compare it with the 
ideal theoretical case. There are different approaches are 
used by the scientific community to calculate the speedup 
\citep{xian10,gustafson88} which is written in the current 
article as
\begin{equation}
	{Sp}(m,N) = \frac{T(m,s)}{T(m,N)} \, \mbox{.}
	\label{eq:speedup}
\end{equation}
in which $T$ stands for the time spent by mesh $m$ to perform 
one thousand iterations, $N$ represents the number of 
computational cores and $s$ is the starting point of the 
scalability study. The strong scaling efficiency of a given 
mesh configuration, as a function of the number of processors, 
is written considering the law of Amdahl \cite{amdahl67} as
\begin{equation}
	\eta(m,N) = \frac{{Sp}(m,N)}{N} \, \mbox{.}
	\label{eq:eff}
\end{equation}

\begin{figure}[htb!]
  \begin{center}
    \includegraphics[width=0.75\textwidth,
    trim=0.25cm 0.25cm 0.25cm 0.25cm,clip]
    {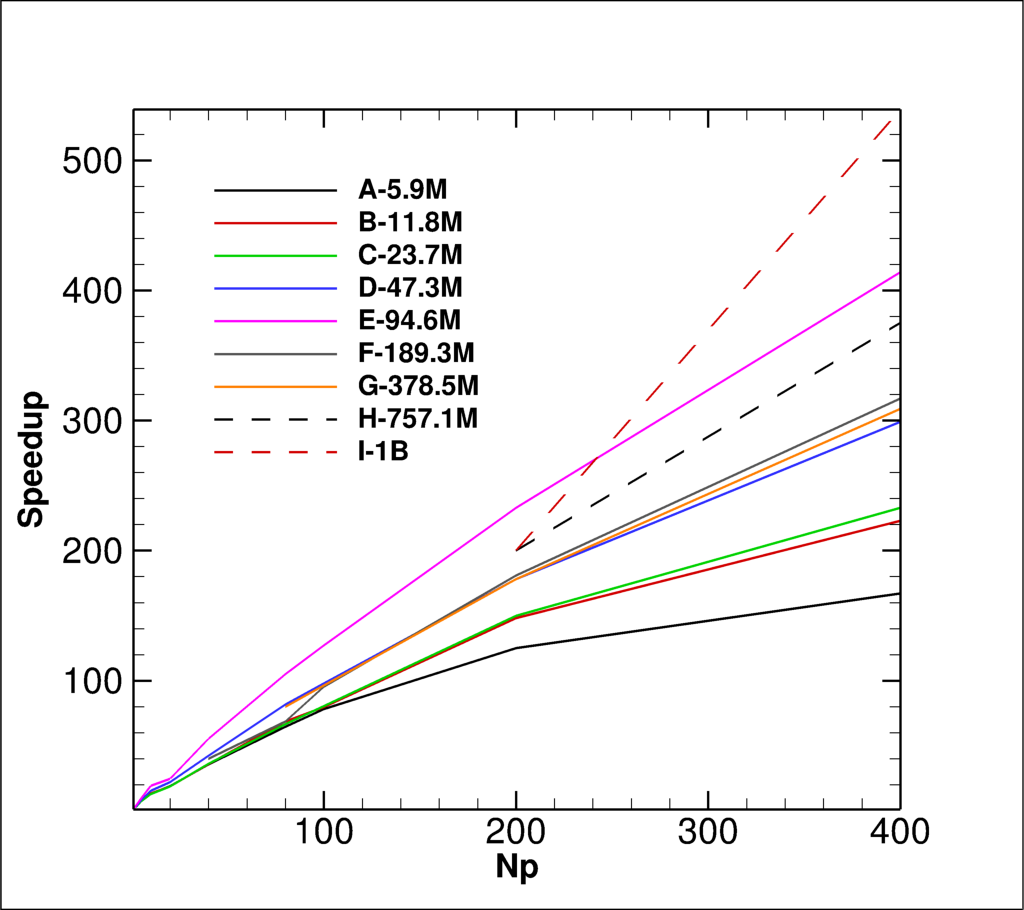}
    \caption{Speedup curve of the LES solver for 
    different mesh sizes.}\label{fig:speedup}
  \end{center}
\end{figure}

\begin{figure}[htb!]
  \begin{center}
    \includegraphics[width=0.75\textwidth,
    trim=0.25cm 0.25cm 0.25cm 0.25cm,clip]
    {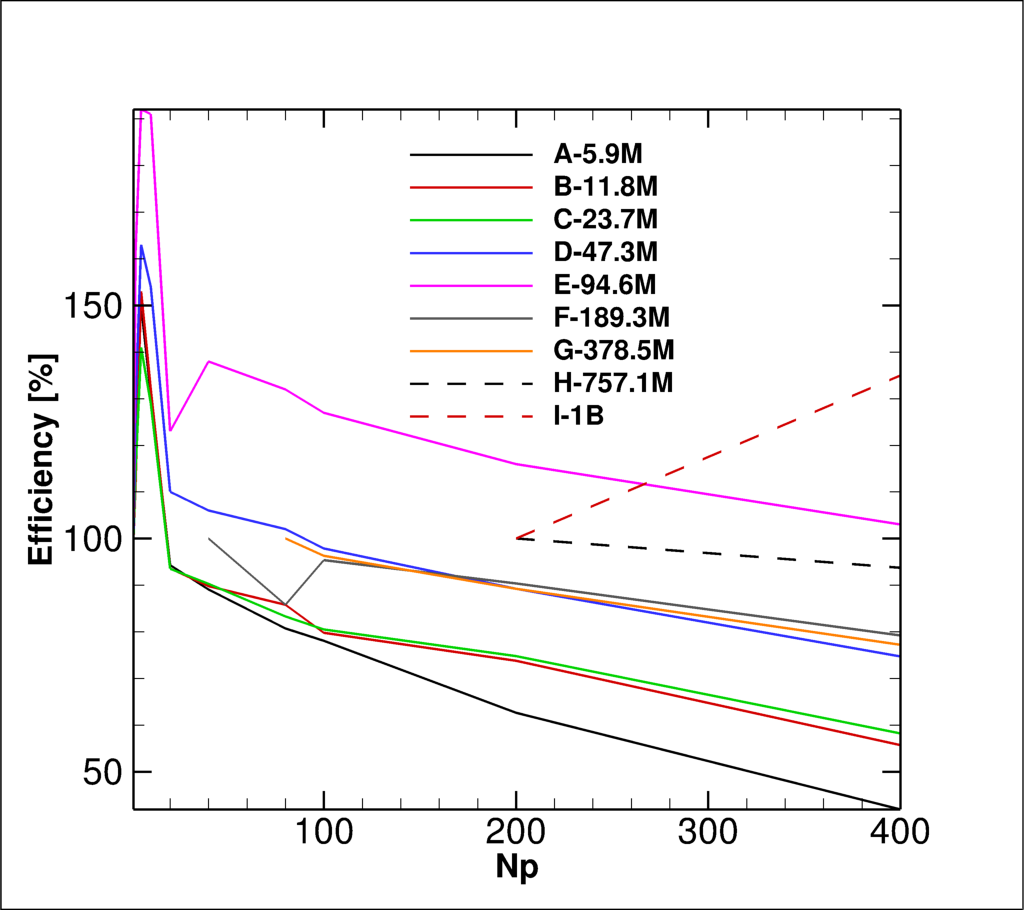}
    \caption{Parallel efficiency of the LES solver for 
    different mesh sizes.}\label{fig:efficiency}
  \end{center}
\end{figure}

More than 300 calculations are performed when considering all 
the partitioning configurations and different meshes evaluated 
in the present paper. The averaged time per iteration is 
calculated for all numerical simulations in order to study the 
scalability of the solver. The evolution of speedup and efficiency, 
as a function of the number of processors, for the nine grids used 
in the current work are presented in Figs.\ \ref{fig:speedup} and 
\ref{fig:efficiency}. The investigation indicates a good scalability 
of the code. Meshes with more than 50 million points present 
efficiency bigger than 75\% when running with 400 computing cores 
in parallel. Moreover, mesh E, which has $\approx$ 95 million degrees 
of freedom, presented an efficiency which equivalent to the ideal 
case, $\approx$ 100\%, when using the maximum number of resources 
evaluated. One can notice a superlinear scalability for the cases 
evaluated in the present article. This behavior can be explained by 
the fact that cache memory can be accessed more efficiently when 
increasing the total number of processors for a given grid 
configuration since more computational resources for the same load 
means less cache miss \citep{Ristov16,Gusev14}.

Increasing the size of a computational problem can generate a better 
scalability study. The time spent with computation becomes more 
significant when compared to the time spent with communication with 
the growth of a problem. One can notice such effect for meshes A, B, 
C, D and E. The speedup and the efficiency increase with the growth of 
the mesh size. However, such scalability improvement does not happen 
from mesh E to meshes F, G, H and I. This behavior is originated because 
the reference used to calculate speedup and efficiency is not the same 
for all grid configurations. The studies performed using meshes F, G, H 
and I does not use the serial computation as a reference, which is not 
the case of calculations performed using mesh A, B, C, D and E.




\section{Compressible Jet Flow Simulation}\label{sec:val}

This section presents a compilation of results achieved from the 
simulation of a supersonic jet flow configuration. This calculation 
was performed in order to validate the LES code, and it is included 
here simply to demonstrate that the numerical tool is indeed capable of 
presenting physically sound results for the problem of interest. 
Results are compared to numerical \citep{Mendez10,Mendez12} and to
experimental data \citep{bridges2008turbulence}. The details of this 
particular simulation are published in the work of 
{\em Junqueira et. al.}\cite{Junior18-abcm}.

\begin{figure}[htb!]
  \begin{center}
    \subfigure[A Two-dimensional surface, colored by velocity 
    magnitude contours, extracted from the full geometry.]{
    \includegraphics[trim= 5mm 5mm 5mm 5mm,clip,width=0.4\textwidth]
	{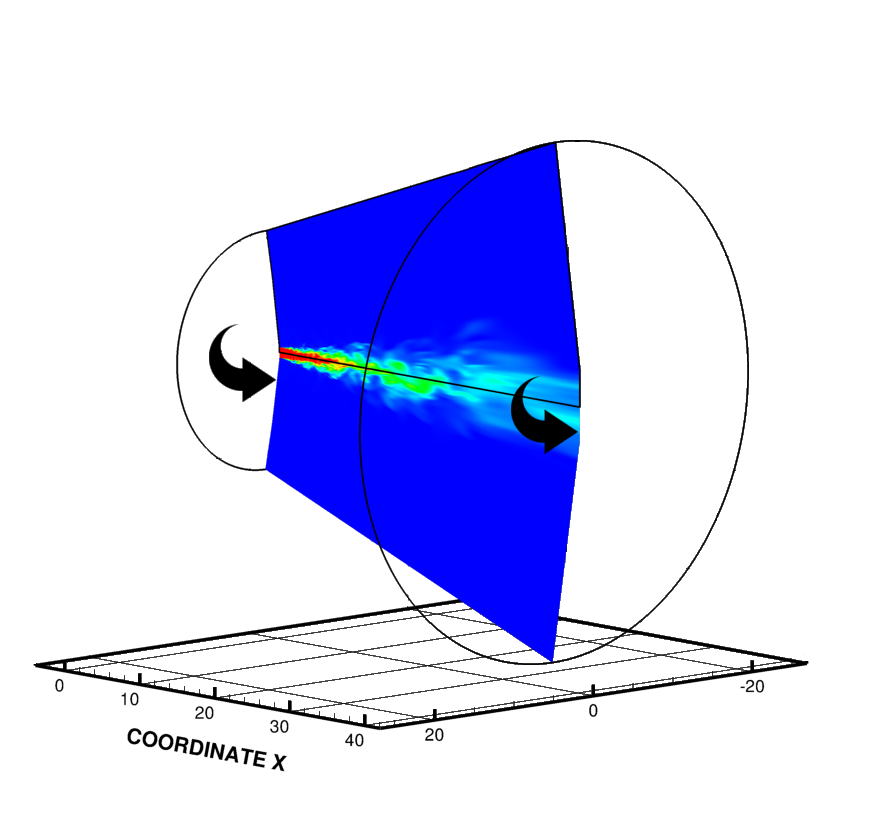} 
    \label{fig:mesh}
    }
	\subfigure[A Two-dimensional surface extracted from the 
    full domain superimposed by grid points distribution.]{
    \includegraphics[trim= 5mm 5mm 5mm 5mm,clip,width=0.45\textwidth]
	{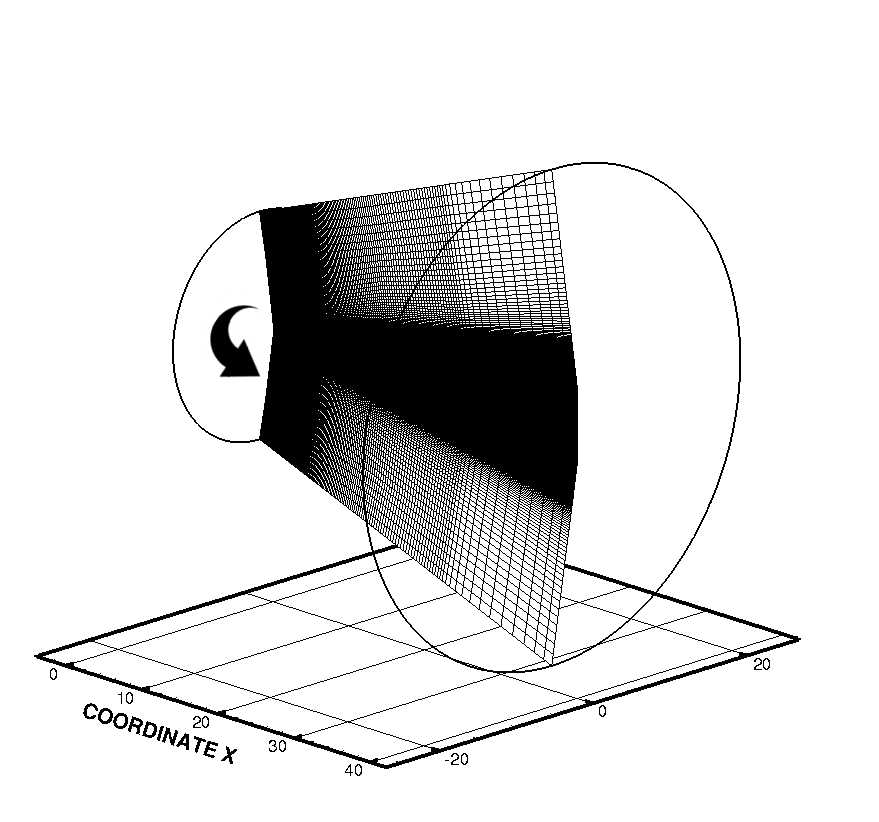}
    \label{fig:geom-val}
    }
	\caption{Illustration of geometry and mesh used in the 
	validation of the LES solver.}\label{fig:mesh-geom}
  \end{center}
\end{figure}

A geometry is created using a divergent shape whose axis length is 
40 times the jet entrance diameter, $D$.  
Figure \ref{fig:mesh-geom} illustrates a 2-D cut of the geometry 
and the grid point distribution used on the validation of the solver. 
The mesh presents approximately 50 million points.
The calculation is performed using 500 computational cores.

An unheated perfectly expanded jet flow is studied for the present validation. 
The jet entrance Mach number is $1.4$\@. The pressure ratio, $PR=P_{j}/P_\infty$, 
and the temperature ratio, $TR=T_{j}/T_\infty$, between the jet entrance and 
the ambient freestream conditions, are equal to one, {\em i.e.}, $PR = 1$ and 
$TR=1$. The $j$ subscript identifies the properties at the jet entrance and the 
$\infty$ subscript stands for properties at the farfield region. The Reynolds 
number of the jet is $Re = 1.57 \times 10^{6}$, based on the jet entrance 
diameter, D\@. The time increment, $\Delta t$, used for the validation study 
is $1\times 10^{-4}$ dimensionless time units. 

\begin{figure}[htb!]
  \centering
  \subfigure[Time averaged axial component of velocity, $\langle U \rangle$.]
  {\includegraphics[trim= 5mm 5mm 5mm 5mm, clip, width=0.45\textwidth]{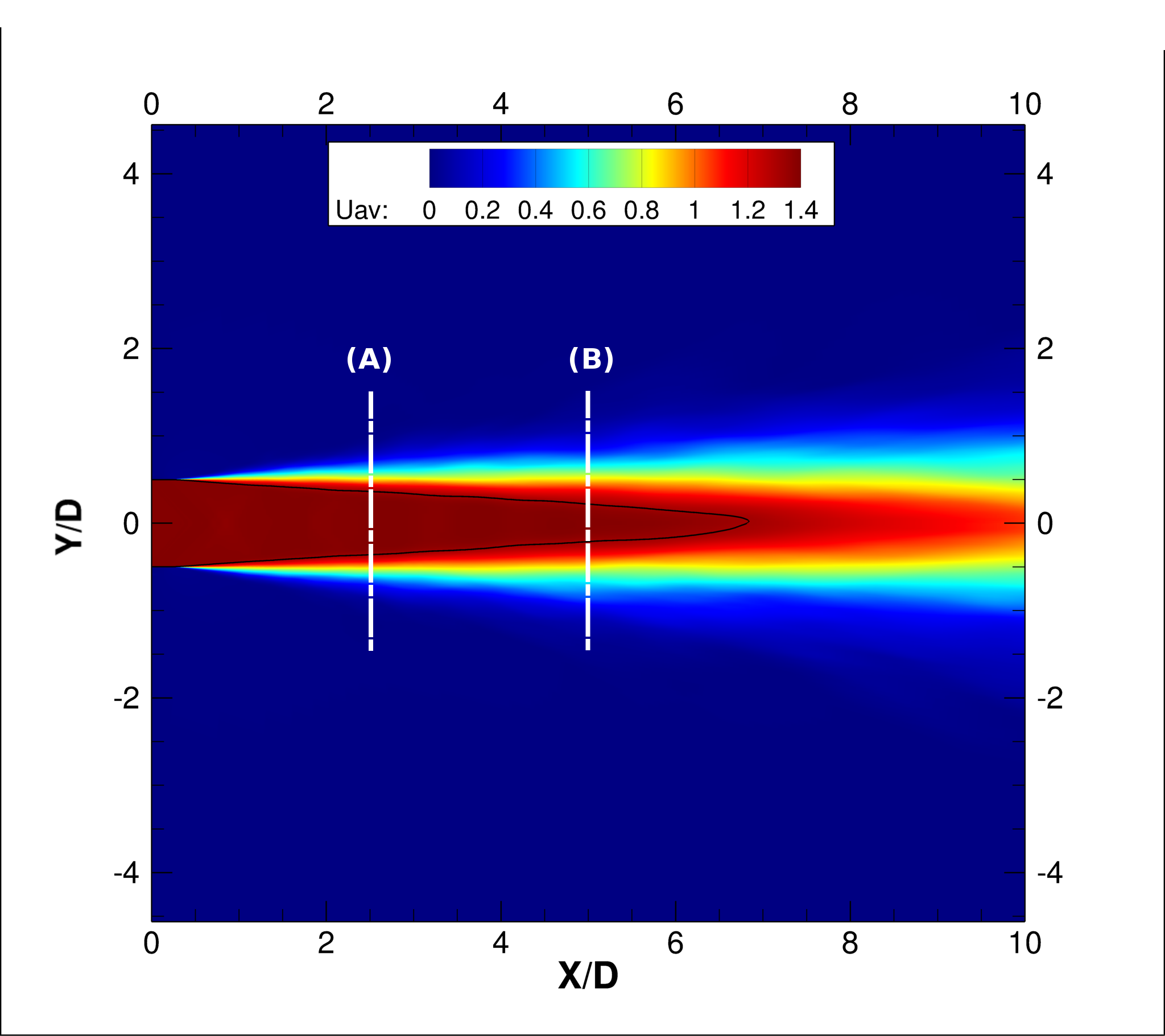}\label{fig:uav}}
  \subfigure[RMS values of the fluctuating part of velocity axial component,
             $u_{RMS}^{*}$.]
    {\includegraphics[trim= 5mm 5mm 5mm 5mm, clip, width=0.45\textwidth]{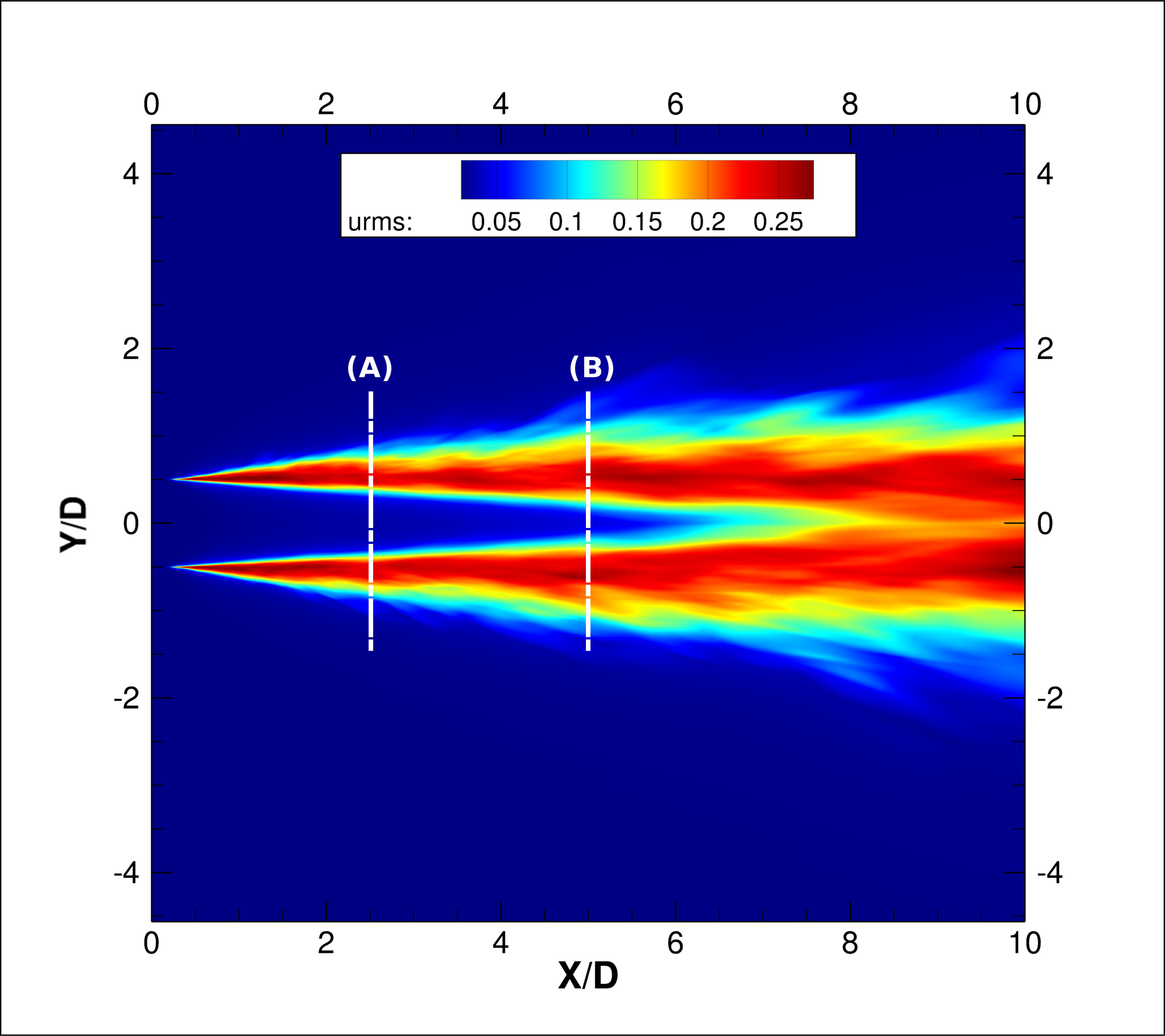}\label{fig:urms}}
	\caption{Lateral view of distributions of $\langle U \rangle$ and $u_{RMS}^{*}$. 
    The white dashed lines indicate the positioning of radial cuts where data are 
    extracted and averaged. The solid black line in {\bf (a)} represents the potential 
    core of the jet.}
	\label{fig:dist-u}
\end{figure}

The boundary conditions previously presented in the {\it Large 
Eddy Simulation Formulation} section, are applied in the current simulation. 
The stagnation state of the flow is set as an initial condition of the 
computation. The calculation runs a predetermined period of time until 
reaching the statistically steady flow condition. This first pre-simulation 
is important to assure that the jet flow is fully developed and turbulent. 
Computations are restarted and run for another period after achieving the 
statistically stationary state flow. Hence, data are extracted and recorded 
in a predetermined frequency. Figure \ref{fig:dist-u} indicates the 
positioning of the two surfaces, (A) and (B), where data are extracted and 
averaged through time. Cuts (A) and (B) are radial profiles at $2.5D$ and 
$5.0D$ units downstream of the jet entrance. Flow quantities are also averaged 
in the azimuthal direction when the radial profiles are calculated. 

Figures \ref{fig:uav} and \ref{fig:urms} present distributions of time averaged axial 
component of velocity and root mean square values of the fluctuating part of the axial 
component of velocity, which are represented in the present work as $\langle U \rangle$ 
and $u_{RMS}^{*}$, respectively. The solid black line indicated in Fig.\ \ref{fig:uav} 
represents the potential core of the jet, which is defined as the region where the 
time averaged axial velocity component is at least $95\%$ of the velocity of the jet 
at the inlet.

\begin{figure}[htb!]
  \centering
  \subfigure[X=2.5D]
    {\includegraphics[trim= 5mm 5mm 5mm 5mm, clip, width=0.45\textwidth]{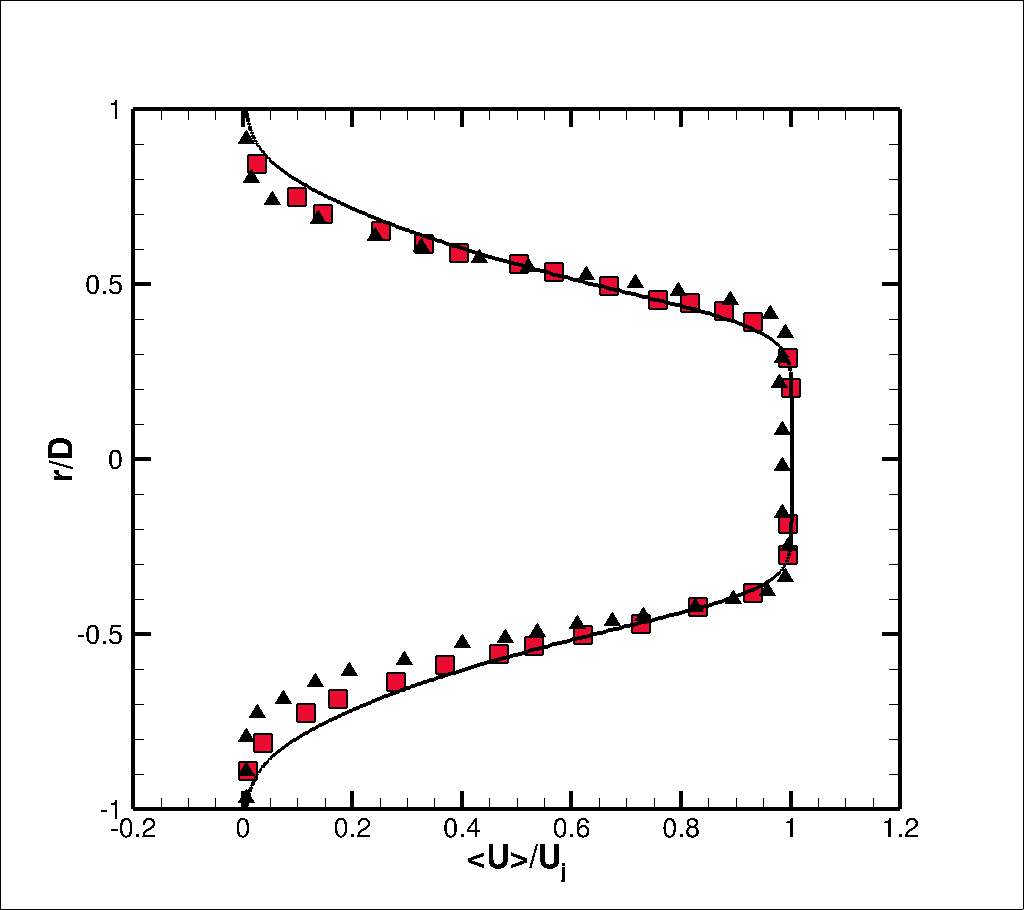}\label{fig:uav-2-5D}}
  \subfigure[X=5.0D]
    {\includegraphics[trim= 5mm 5mm 5mm 5mm, clip, width=0.45\textwidth]{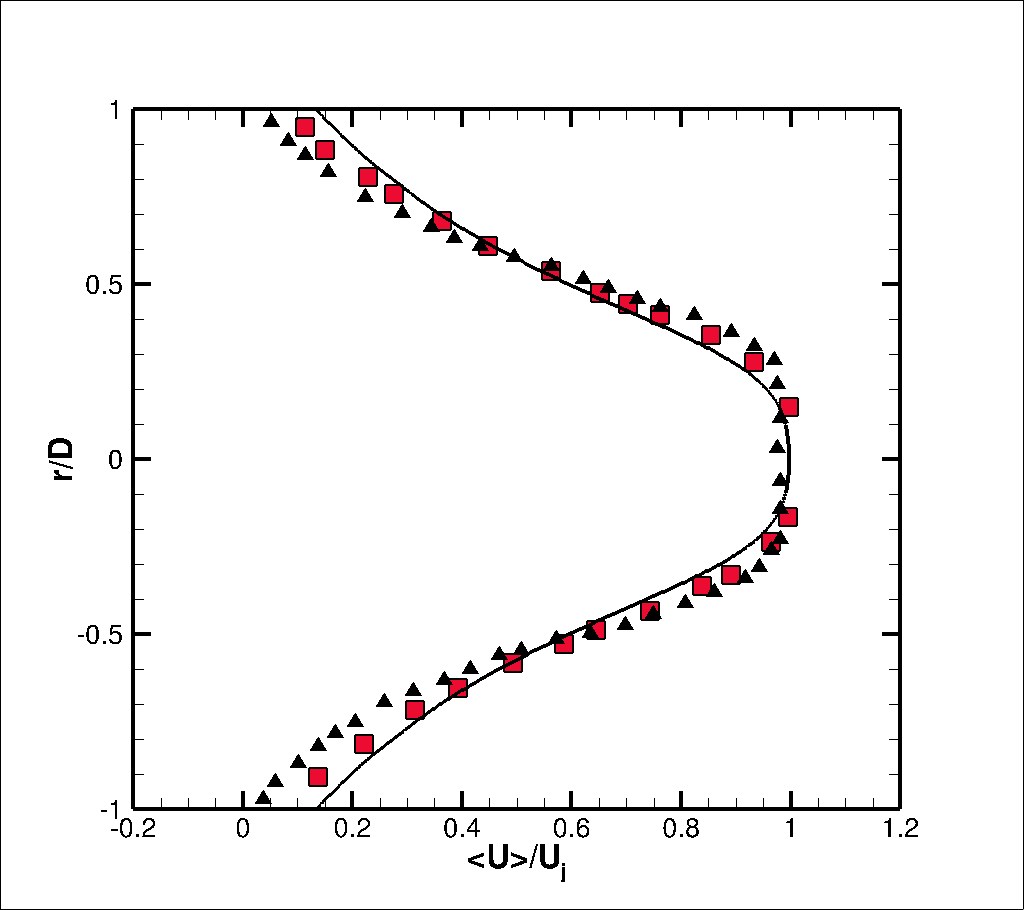}\label{fig:uav-5-0D}}
	\caption{Profiles of the averaged axial component of velocity, 
    $\langle U \rangle$, at $2.5D$ and $5.0D$ from the entrance: 
	(\textbf{--}) JAZzY results; (${\color{red}\blacksquare}$) numerical data \citep{Mendez10,Mendez12}; 
    (${\color{black}\blacktriangle}$) experimental data \citep{bridges2008turbulence}.}
	\label{fig:prof-uav}
\end{figure}

\begin{figure}[htb!]
  \centering
  \subfigure[X=2.5D]
  {\includegraphics[trim= 5mm 5mm 5mm 5mm, clip, width=0.45\textwidth]{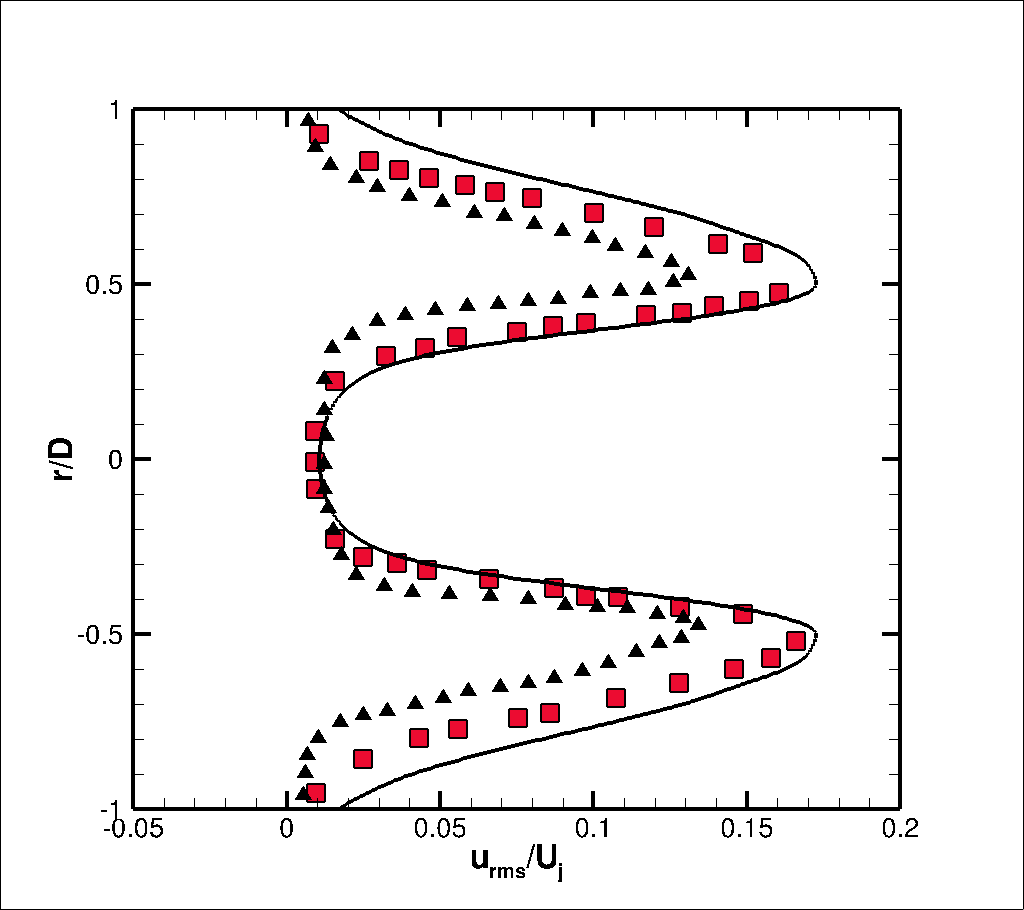}\label{fig:urms-2-5D}}
  \subfigure[X=5.0D]
    {\includegraphics[trim= 5mm 5mm 5mm 5mm, clip, width=0.45\textwidth]{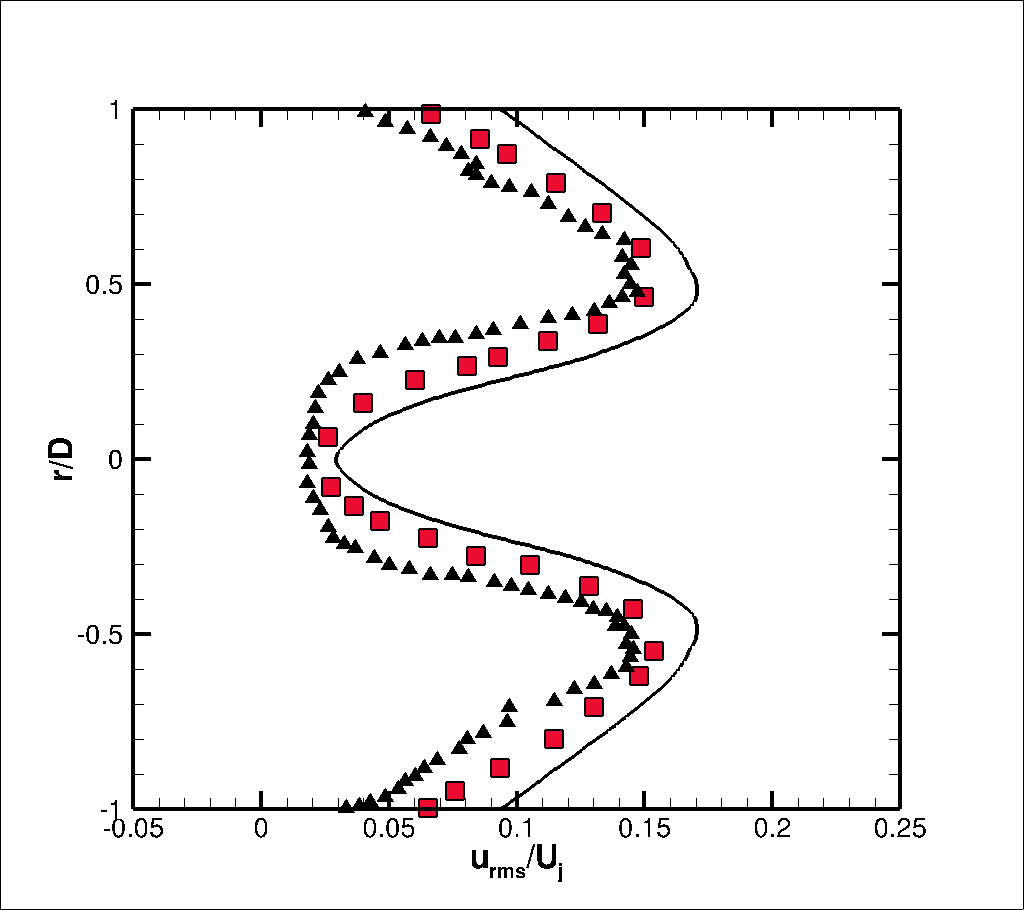}\label{fig:urms-5-0D}}
	\caption{Profiles of the RMS of the fluctuation part of axial component of 
    velocity, $u_{RMS}^{*}$, at $2.5D$ and $5.0D$ from the entrance: 
	(\textbf{--}) JAZzY results; (${\color{red}\blacksquare}$) numerical data \citep{Mendez10,Mendez12}; 
    (${\color{black}\blacktriangle}$) experimental data \citep{bridges2008turbulence}.}
	\label{fig:prof-urms}
\end{figure}

Dimensionless profiles of $\langle U \rangle$ and $u^{*}_{RMS}$ at the cuts along 
the mainstream direction of the computational domain are compared with numerical 
and experimental results in Figs.\ \ref{fig:prof-uav} and 
\ref{fig:prof-urms}, 
respectively. The solid line stands for results achieved using the JAZzY code while 
square and triangular symbols represent numerical \citep{Mendez10,Mendez12} 
and experimental \citep{bridges2008turbulence} data, respectively. 
The averaged profiles obtained in the present work correlate well with the reference 
data at the two positions compared here.  
It is important to remark that the LES tool 
can provide good predictions of supersonic jet flow configurations when using a 
sufficiently fine grid point distribution. Therefore, efficient massive parallel 
computing is mandatory in order to achieve good results. 

\begin{figure}[htb!]
  \centering
    {
	  \includegraphics[trim= 1mm 1mm 1mm 1mm, clip, width=0.45\textwidth]
      {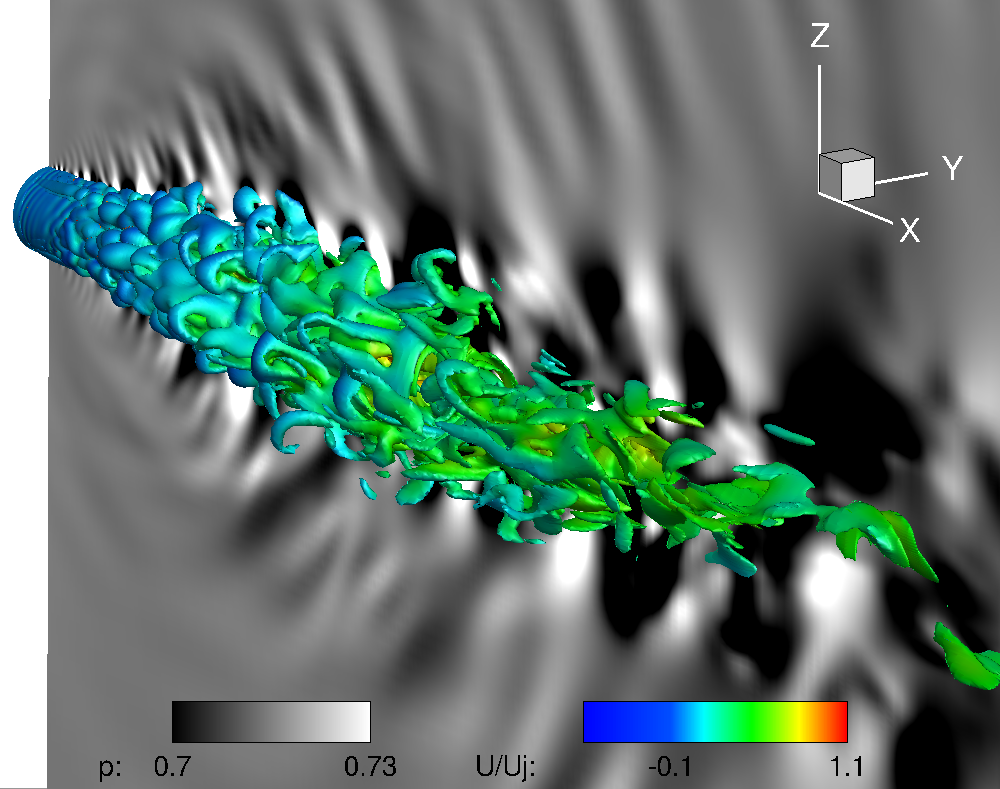}
    }
	\caption{Instantaneous lateral view of pressure contours, in greyscale, 
    superimposed by 3-D velocity magnitude contours, in color.}
	\label{fig:3d-vel}
\end{figure}

\begin{figure}[htb!]
  \centering
    {\includegraphics[trim= 1mm 1mm 1mm 1mm, clip, width=0.9\textwidth]{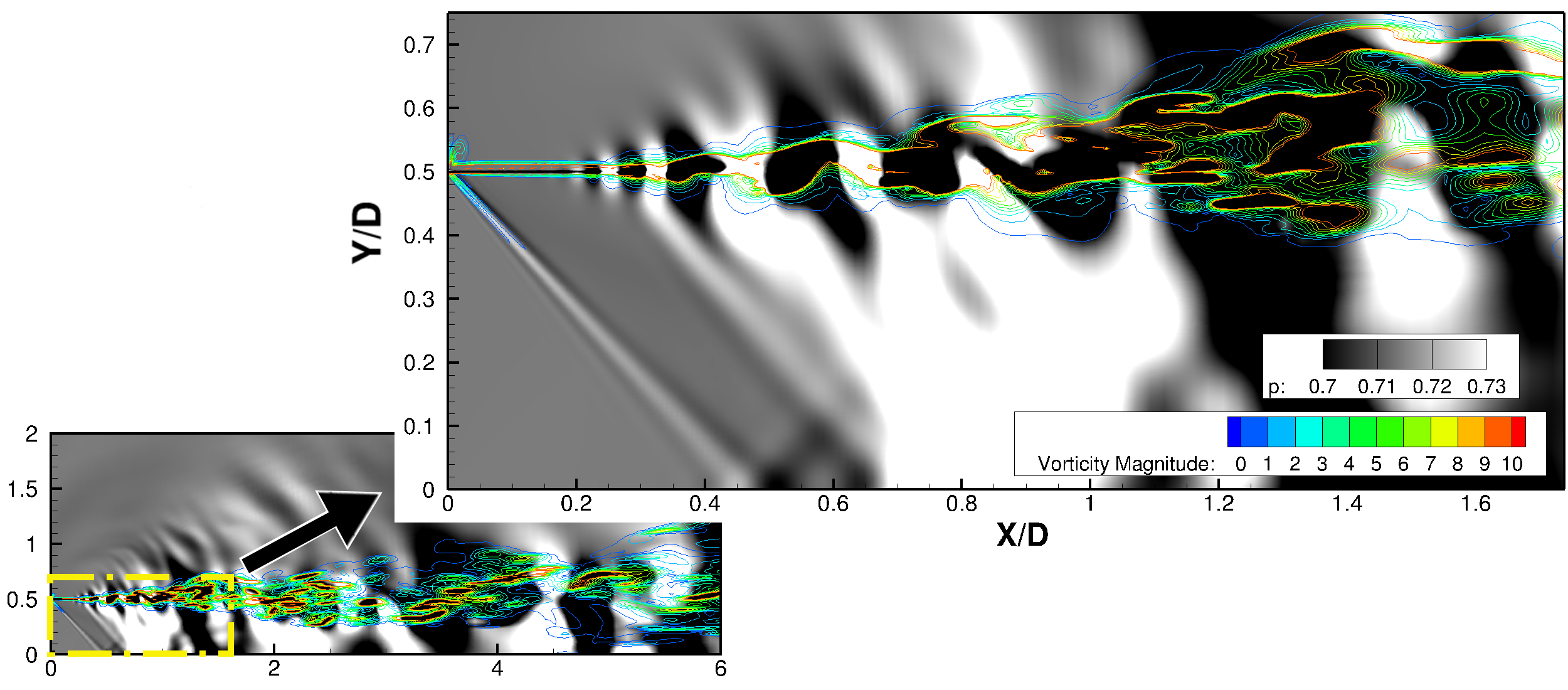}\label{fig:vort-zoom}}
	\caption{Lateral and detailed view of pressure contours, in greyscale,
    superimposed by vorticity magnitude contours, in color. The yellow box 
	indicates the region illustrated in the detailed view.}
	\label{fig:press-vort}
\end{figure}

Figures \ref{fig:3d-vel} and \ref{fig:press-vort} present a lateral 
view of an instantaneous visualization of the pressure contours, in greyscale, 
superimposed by 3-D velocity magnitude countours and vorticity magnitude contours 
respectively, in color, calculated by the LES tool discussed in the present paper. 
A detailed visualization of the region indicated in yellow, at the 
jet entrance, is shown in Fig.\ \ref{fig:press-vort}. The resolution of flow features 
obtained from the jet simulation is more evident in this detailed plot of the jet 
entrance. One can clearly notice the compression waves generated at the shear layer, 
and their reflections at the jet axis. Such resolution is important to observe 
details and behavior of such flow configuration in order to understand the acoustic 
phenomena which is presnt in supersonic jet flow configurations.



\section{Concluding Remarks}

The current work is concerned with the performance of a 
computational fluid dynamics tool for aeroacoustics 
applications when using a national supercomputer. The HPC 
system, Euler, from the University of S\~{a}o Paulo presents 
more than 3000 computational cores and a maximum theoretical 
peak of 127.4 TFLOPS. The numerical solver is developed by 
the authors to study supersonic jet flows. Simulations of such 
flow configurations are expensive and need efficient parallel 
computing. Therefore, strong scalability studies of the solver 
are performed on the Euler supercomputer in order to evaluate 
if the numerical tool is capable of efficiently using 
computational resources in parallel.

The computational fluid dynamics solver is developed using 
the large eddy simulation formulation for perfectly expanded 
supersonic jet flow. The equations are written using a 
finite-difference centered  
spatial discretization with the addition of artificial 
dissipation. The time integration is performed using a 
five-steps 
Runge-Kutta scheme. Parallel computing is achieved through 
non-blocking message passing interface protocols and 
inter-partition data are allocated using ghost points. Each 
MPI partition reads and writes its own portion of the mesh that 
is created on pre-processing routine.

A geometry and a flow condition are defined for the scalability 
study performed in the present work. Nine point distributions and 
different partitioning configurations are used in order to 
evaluate the parallel code under different workloads. The size of 
the grid configurations start with 5.9 million points and rise up 
to approximately 1.0 billion points. Calculations perform 1000 
iterations or 24 hours of computation using up to 400 cores in 
parallel. The CPU time per iteration is averaged when the 
simulation is finished in order to calculate the speedup and 
scaling efficiency. More than 300 simulations are performed for 
the scalability study when considering different workloads and 
partitioning configurations.  

The code presented a good scalability for the calculations run 
in the current paper. The averaged CPU time per iteration decays 
with the increase in the number of processors in parallel for all 
computation performed by the large eddy simulation solver evaluated 
in the present work. Meshes with more than 50 million points 
indicated an efficiency greater than 75\%. The problem with 
approximately 100 million points presented speedup of 400 and 
efficiency of 100\% when running on 400 computational cores in 
parallel. Such performance is equivalent to theoretical behavior in 
parallel. It is important to remark the ability of the parallel 
solver to treat very dense meshes as the one tested in the present 
paper with approximately 1.0 billion points. Large eddy simulation 
demands very refined grids in order to have a good representation of 
the physical problem of interest. Therefore, it is important to 
perform simulations of such configuration with a good computation 
efficiency and the present scalability study article can all be seen 
as a guide for future simulations using the same numerical tool on 
the Euler supercomputer.



\section*{Acknowledgments}
\label{acknowledgments}

The authors gratefully acknowledge the partial support for this 
research provided by Conselho Nacional de Desenvolvimento Cient\'{i}fico 
e Tecnol\'{o}gico, CNPq, under the Research Grant nos. 309985/2013-7, 
400844/2014-1 and 443839/2014-0. The authors are also indebted to the 
partial financial support received from Funda\c{c}\~{a}o de Amparo 
\`{a} Pesquisa do Estado de S\~{a}o Paulo, FAPESP, under the Research 
Grant nos. 2013/07375-0 and 2013/21535-0. 

\section*{Conflict of Interest}

The authors declare that they have no conflict of interest.


\bibliographystyle{plain}
\small{
\bibliography{main}
}

\end{document}